\documentclass[aps, pre, twocolumn, 10pt]{revtex4-2}
\usepackage{amsfonts,amsmath,bm,amssymb,color}
\usepackage{graphicx}
\usepackage{multirow}
\usepackage{xcolor}
\usepackage{tcolorbox, enumitem}
\usepackage[colorlinks=true, citecolor=blue, linkcolor=blue]{hyperref}
\DeclareMathOperator*{\argmax}{argmax}

\begin{document}

\title{Optimal navigation in two-dimensional flows:\\ Control theory and reinforcement learning}

\author{Vladimir Parfenyev}\email{parfenius@gmail.com}
\affiliation{Landau Institute for Theoretical Physics, Russian Academy of Sciences, 1-A Akademika Semenova av., 142432 Chernogolovka, Russia}
\affiliation{HSE University, Faculty of Physics, Myasnitskaya 20, 101000 Moscow, Russia}

\date{\today}

\begin{abstract}
Zermelo's navigation problem seeks the trajectory of minimal travel time between two points in a fluid flow. We address this problem for an agent -- such as a floating drone or active particle -- that is advected by a two-dimensional flow, self-propels at a fixed speed smaller than or comparable to the characteristic flow velocity, and can steer its direction. The flows considered span increasing levels of complexity, from steady solid-body rotation and time-dependent sink-vortex to the Taylor-Green flow and turbulence in the inverse energy cascade regime. Although optimal-control theory provides time-minimizing trajectories, these solutions become unstable in chaotic regimes characterized by positive finite-time Lyapunov exponents. To design robust navigation strategies, we apply reinforcement learning and compare Q-learning with a one-step actor-critic algorithm. Both methods achieve successful navigation, yielding mean travel times within 3-10\% of optimal-control solutions in regular flows, while the discrepancy increases to 35–75\% in time-dependent turbulent flows. Finally, we show that agents trained on coarse-grained turbulent flows generalize to the full velocity field. This robustness to incomplete flow information is essential for practical navigation in real-world oceanic and atmospheric environments.
\end{abstract}

\maketitle

\section{Introduction}

The quest for the optimal path between two points, a cornerstone of navigation and control theory, finds profound expression in the classical Zermelo's navigation problem. Originally formulated by Ernst Zermelo in his seminal work~\cite{zermelo1931navigationsproblem}, the problem seeks the trajectory of minimal travel time for an agent moving with constant self-propulsion speed in a known fluid flow by adjusting the steering direction. Beyond its theoretical elegance, this framework holds critical importance for a variety of modern applications, from micro-robots navigating vascular networks for targeted drug delivery~\cite{nelson2010microrobots, park2017multifunctional, manzari2021targeted}, to autonomous underwater and aerial vehicles operating in oceanic and atmospheric environments~\cite{trincavelli2008towards, lermusiaux2017future, bellemare2020autonomous, chai2020monitoring}. The core challenge lies in the interplay between the agent's own propulsion and the advective power of the flow, which often dominates the dynamics, necessitating a control strategy that can exploit environmental forces rather than simply resist them.

The study of Zermelo's problem has traditionally been approached through the lens of optimal control theory. Within this framework, one can derive a differential equation governing the evolution of the agent's steering direction, which serves as the necessary condition for optimality (see, e.g., Ref.~\cite[Sec. 2.7]{bryson2018applied}). For relatively simple flow fields, this equation admits analytical solutions~\cite{techy2011optimal, liebchen2019optimal}, enabling the construction of optimal trajectories. In more complex environments, however, numerical techniques are required. The challenge is further complicated by the absence of a known initial condition for the steering equation, which in practice must be determined using shooting methods~\cite{kirk2004optimal, biferale2019zermelo}. Recently, this approach was generalized to the curved manifolds for steady environments~\cite{piro2021optimal, piro2022efficiency}.

This classical formulation has a crucial limitation. The degree of freedom associated with the steering direction makes the system susceptible to chaotic behavior and the resulting optimal trajectories can be unstable even in regular external flows~\cite{biferale2019zermelo, piro2021optimal, piro2022efficiency}. In these cases, infinitesimal uncertainties in initial conditions or control inputs are exponentially amplified, leading to drastic deviations from the planned route~\cite{angelo2009chaos}. Moreover, there is no longer a systematic way to determine a globally optimal solution to Zermelo's problem. At best, one can compute several locally optimal trajectories and then select the one with the minimal travel time.

A complementary approach capable of producing robust navigation strategies is based on reinforcement learning (RL). In this framework, the agent learns through interaction with the environment~\cite{sutton1998reinforcement}. Early studies have demonstrated the potential of this approach. For instance, Q-learning was used in Refs.~\cite{yoo2016path, schneider2019optimal} for path planning in steady flows, while Refs.~\cite{biferale2019zermelo, buzzicotti2020optimal} explored a one-step actor-critic algorithm applied to Zermelo's problem in two-dimensional turbulent flows. A more advanced deep RL method A2C was later applied in Ref.~\cite{nasiri2022reinforcement}; however, the analysis was focused on steady currents. Such environments are considerably simpler than time-dependent ones, and heuristic strategies~\cite{piro2022optimal} can be employed to correct deviations from the optimal-control path caused by instabilities or noise. We emphasize that in Zermelo's problem the agent has access to the global velocity field, in contrast to widely studied navigation problems where decisions rely on local sensory input~\cite{colabrese2017flow, gustavsson2017finding, alageshan2020machine, yang2020efficient, gunnarson2021learning, monthiller2022surfing, calascibetta2023optimal, mecanna2025critical}. For a comprehensive overview of the literature, we refer the reader to Ref.~\cite{nasiri2023optimal}.
 
Here, we extend recent studies in three main ways. First, much of the literature has considered applications of Zermelo's equation to navigation in steady currents (with a notable exception in Ref.~\cite{techy2011optimal}), and most optimal-control studies~\cite{biferale2019zermelo, piro2022efficiency, piro2021optimal, piro2022optimal, liebchen2019optimal, schneider2019optimal} have focused exclusively on steady environments. We complement these efforts by computing optimal-control solutions in unsteady flows, including two-dimensional turbulence. Although these trajectories may be unstable and represent only local optima, they nevertheless provide upper bounds for the minimum travel time and serve as useful benchmarks for evaluating the performance of more robust algorithms.

Second, we implement both Q-learning and a one-step actor-critic algorithm and compare their performance with each other and with the optimal-control solutions across a wide range of two-dimensional flows. These flows include steady solid-body rotation, a time-varying sink-vortex, the Taylor-Green flow, and both snapshot and time-dependent turbulent fields. Our focus on two-dimensional environments is motivated by their computational tractability and their relevance to large-scale atmospheric and oceanic currents~\cite{vallis2017atmospheric, boffetta2012two}. The analysis shows that in all scenarios both RL methods learn strategies that reach the target. However, their travel times exceed the optimal-control solutions by up to 10\% in regular flows and up to 35–75\% in time-varying turbulence. These results indicate that tabular RL methods employed here struggle to identify effective navigation strategies in complex environments, leaving significant room for improvement through more advanced deep RL approaches~\cite{nasiri2022reinforcement, mecanna2025critical}, particularly in time-dependent turbulent settings.

Third, previous studies have examined the effect of Gaussian white noise on navigation, which is relevant for small-scale active particles and models thermal fluctuations~\cite{schneider2019optimal, muinos2021reinforcement, piro2022optimal, piro2022efficiency}. For agents operating in turbulent environments, however, a different source of uncertainty arises: operational forecasts are inherently coarse-grained, forcing agents to operate with incomplete information about the external flow. To address this issue, we train RL agents on coarse-grained turbulent flows and then evaluate their performance in fully resolved environments. By varying the coarse-graining threshold, we control the amount of flow information available to the agent. The results indicate that RL methods achieve successful navigation even when small-scale flow information is missing.


\section{Problem Setup}\label{sec:2}

We study the motion of an agent navigating a two-dimensional flow field $\bm v(\bm r, t)$, which is governed by the kinematic equation
\begin{equation}\label{eq:motion}
\dfrac{d \bm R_t}{dt} = \bm v(\bm R_t, t) + V_0
\begin{pmatrix}
\cos \theta_t \\
\sin \theta_t
\end{pmatrix},
\end{equation}
where $\bm R_t = (X_t, Y_t)$ is the position of the agent at time $t$ in Cartesian coordinates. The dynamics combines passive advection by the background flow $\bm v (\bm R_t, t)$, which represents the external environmental velocity such as oceanic or atmospheric currents, and active self-propulsion at constant speed $V_0$. The steering direction $\theta_t$ serves as the control variable that the agent can adjust in real-time. The objective of Zermelo's problem~\cite{zermelo1931navigationsproblem} is to find a control policy $\theta_t = \theta(\bm R_t, t)$ that minimizes the total travel time $T$ for a journey between two predefined points $\bm r_A$ and $\bm r_B$. 

The nature of the solution is dependent on the relative strength of the flow~\cite{biferale2019zermelo}. In the regime, where the agent's speed $V_0$ significantly exceeds the characteristic flow velocity, one may expect that a trivial strategy of continuously orienting the propulsion vector directly toward the target $\bm r_B$ is close to optimal. The problem becomes more challenging in the opposing regime, where the flow exerts a strong influence. Under these conditions, the agent must strategically ride the current to exploit favorable flow structures for an overall reduction in travel time. This latter regime is the primary focus of our work.

\section{Background Flows}\label{sec:3}

This section introduces background flows considered in our study. The first two examples are adopted from Ref.~\cite{techy2011optimal}, and for them the Zermelo's problem admits an analytical solution. 

\textit{Steady vortex.} In the first example, the background velocity corresponds to steady solid-body rotation
\begin{equation}\label{eq:vortex}
    \bm v (x,y) = \left( -\omega y, \; \omega x \right)^T.
\end{equation}
The starting point is $(x_A, y_A) = \left(1/2, \sqrt{3}/2 \right)$ and our aim is to reach the ending point $(x_B, y_B) = (1, 0)$ in the shortest time. The agent's speed is $V_0=1$, and we set $\omega = 0.9$. 

\textit{Time-varying sink-vortex.} In the second example, the background velocity is given by
\begin{equation}\label{eq:sink+rotation}
    \bm v (x,y, t) =
    \begin{pmatrix}
    -0.3 x - (0.5-t)y \\
    -0.3 y + (0.5-t)x
    \end{pmatrix}.
\end{equation}
The flow model represents a constant sink plus a vortex that linearly changes vorticity with time. The starting point is $(x_A, y_A) = \left(1/2, \sqrt{3}/2 \right)$, the ending point is $(x_B, y_B) = (1, 0)$, and the agent's speed is $V_0=1$. 

\textit{Taylor-Green flow.} In the third example, the velocity field forms a checkerboard pattern of counter-rotating vortices
\begin{equation}\label{eq:TG-flow}
    \bm v (x,y) =
    \begin{pmatrix}
    u_0 \sin(kx)\cos(ky) \\
    -u_0 \cos(kx)\sin(ky)
    \end{pmatrix}.
\end{equation}
We set $u_0=1$, $k=3$, and the agent's speed is $V_0 = 0.1$. The starting point is $(x_A, y_A) = \left(2\pi/3, \pi/3 \right)$ and the ending point is $(x_B, y_B) = (3 \pi/2, 3\pi/2)$. The main challenge in this example is to escape vortex traps along the way. Note that a similar velocity field can be generated on the fluid surface due to the nonlinear interaction of two orthogonal standing waves~\cite{filatov2016nonlinear, filatov2016generation, francois2017wave, xia2019generation, parfenyev2019formation, abella2020measurement}.

\textit{2D Turbulence.} Finally, we consider a two-dimensional incompressible turbulent flow obtained by numerically integrating the 2D Navier-Stokes equations
\begin{equation}\label{eq:NS-equation}
\partial_t \bm v + (\bm v \cdot \nabla) \bm v = - \nabla p + \nu \nabla^2 \bm v - \alpha \bm v + \bm f,
\end{equation}
using the GeophysicalFlows.jl pseudospectral solver~\cite{GeophysicalFlowsJOSS}. The simulation is performed in a doubly periodic square domain of size $L=2\pi$. Here, $\bm v$ is 2D velocity, $p$ is the pressure, $\nu = 5 \cdot 10^{-3}$ is the kinematic viscosity, and $\alpha=0.2$ is the linear bottom drag. The system is driven by an external random forcing $\bm f$ acting at wavenumber $k_f = 5.5$, which injects power $\epsilon = 10$. Further details are provided in the Supplementary Material (SM)~\cite{SM}.

\begin{figure}[t]
    \centering
    \includegraphics[width=0.7\linewidth]{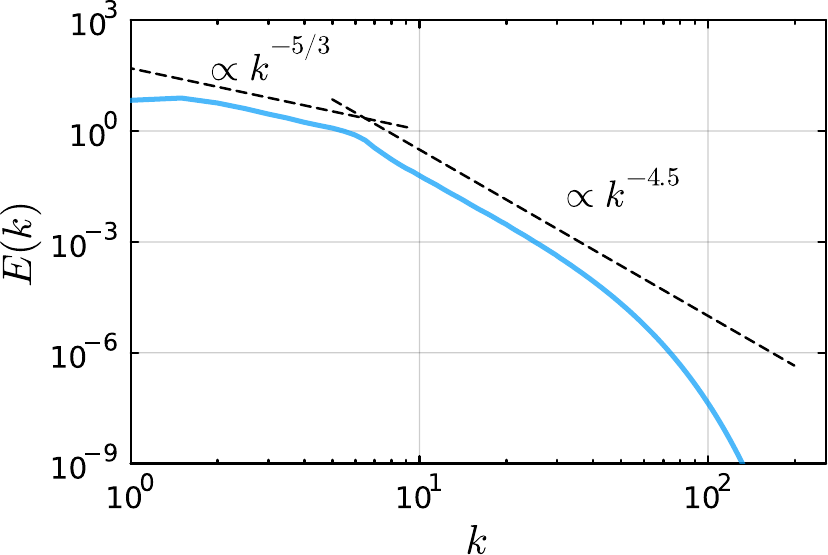}
    \caption{Energy spectrum for DNS of 2D turbulence.}
    \label{fig:1}
\end{figure}

The resulting flow is characterized by dimensionless numbers $Re = UL/\nu \approx 8 \cdot 10^3$ and $Rh = U/(\alpha L) \approx 5$, where $U \approx 6.3$ is the root-mean-square velocity. The mean energy spectrum, shown in Fig.~\ref{fig:1}, confirms the established phenomenology of forced 2D turbulence. In the inverse energy cascade range $k<k_f$, the spectrum follows the theoretical $E(k) \propto k^{-5/3}$ scaling~\cite{boffetta2012two}, while in the direct enstrophy cascade range $k>k_f$, the bottom drag leads to a steeper slope than the classical $E(k) \propto k^{-3}$ prediction~\cite{valadao2025spectrum}. 

To evaluate navigation algorithms, we employ both instantaneous snapshot and the full time-dependent flow data. The agent's self-propulsion speed is set to $V_0 \approx 4.5$, which corresponds to $30\%$ of the maximum flow velocity. For each example, we also consider two different sets of starting and ending points; the corresponding problems are labeled P1 and P2. Experimentally, flows with similar statistical properties can be excited by Faraday waves on the fluid surface~\cite{von2011double, francois2013inverse, francois2014three, colombi2021three, colombi2022coexistence} or by the Lorentz force in a thin conducting fluid layer over a magnet array~\cite{sommeria1986experimental, paret1998intermittency, xia2009spectrally, bardoczi2014experimental, orlov2018large, zhu2024flow}.

\section{Optimal Control Theory}\label{sec:4}

To solve the Zermelo's navigation problem, the equation of motion (\ref{eq:motion}) must be supplemented with an equation describing the evolution of the agent’s heading direction $\theta_t$. One can prove that if an optimal trajectory connecting $\bm r_A$ and $\bm r_B$ exists, then the optimal steering angle must satisfy~\cite{techy2011optimal, SM}
\begin{equation}\label{eq:angle}
\begin{split}
    \dot{\theta}_t = \partial_x v_y \sin^2 \theta_t - \partial_y v_x \cos^2 \theta_t \\
    + (\partial_x v_x-\partial_y v_y) \sin \theta_t \cos \theta_t,
\end{split}
\end{equation} 
where the velocity gradients are evaluated along the agent's trajectory $\bm R_t$. The resulting system must be supplemented with boundary conditions that fix the start and end positions, $\bm R_0 = \bm r_A$ and $\bm R_T = \bm r_B$. Note that the initial heading $\theta_0$ and travel time $T$ are not specified and therefore remain free parameters. 

The standard approach to solve this two-point boundary value problem is the shooting method~\cite{kirk2004optimal, biferale2019zermelo}: an initial guess for $\theta_0$ is refined until the resulting trajectory, obtained by forward integration of the system, satisfies the terminal condition $\bm R_T = \bm r_B$ within a desired tolerance. We emphasize that the considered system may admit multiple solutions: Zermelo's equation (\ref{eq:angle}) expresses only a condition for a local extremum and does not guarantee a global optimum. Note also that once the initial heading $\theta_0$ is known, the optimal path can be integrated autonomously, provided the agent can measure the local velocity field gradients.

\begin{figure*}[t]
\includegraphics[width=0.9\linewidth]{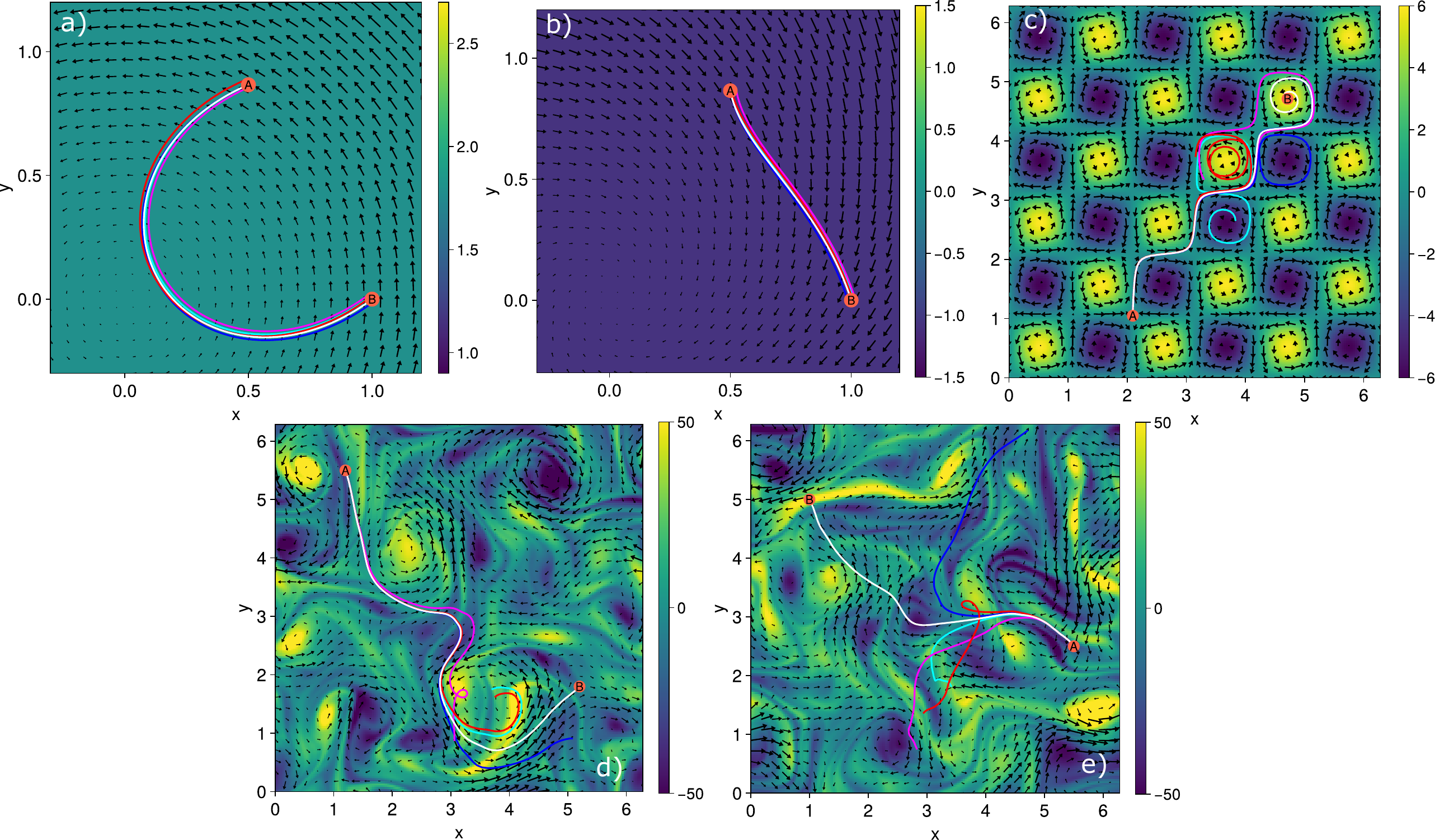}
\caption{Optimal-control trajectories (white lines) for (a) steady vortex, (b) time-varying sink-vortex, (c) Taylor-Green flow, (d) turbulent snapshot (P1), and (e) time-dependent turbulent flow (P1). Colored lines represent perturbed trajectories, illustrating the robustness (a-b) or instability (c-e) of the obtained solutions.}\label{fig:2}
\end{figure*}

Next, we apply the approach to the examples introduced above. For the steady vortex flow (\ref{eq:vortex}), the Zermelo's equation (\ref{eq:angle}) has a solution $\theta_t = \theta_0 + \omega t$. Integrating the equation of motion (\ref{eq:motion}) from the starting point $\bm R_0 = \bm r_A$ yields the trajectory:
\begin{eqnarray}
    \label{eq:VFT1}
    X_t = &x_A \cos(\omega t) + V_0 t\cos(\omega t) \cos \theta_0\\
    \nonumber
    &- y_A \sin(\omega t) - V_0 t \sin(\omega t) \sin \theta_0,\\
    \label{eq:VFT2}
    Y_t = &y_A \cos(\omega t) + V_0 t \cos(\omega t) \sin \theta_0\\
    \nonumber
    &+ x_A \sin(\omega t) + V_0 t \sin (\omega t) \cos \theta_0.&
\end{eqnarray}
The initial heading $\theta_0$ and total travel time $T$ are determined by enforcing the terminal condition $\bm R_T = \bm r_B$. For the specified numerical parameters, we obtain $\theta_0 \approx -1.94$ and $T \approx 1.97$. The found optimal trajectory is illustrated in Fig.~\ref{fig:2}a, which also includes trajectories corresponding to small deviations in the initial heading $\delta \theta_0 = \pi/360$ and position $\Delta = 0.025$. These trajectories stay close to the optimal path on the considered time scale and successfully reach a narrow neighborhood of the target. Based on expressions (\ref{eq:VFT1}) and (\ref{eq:VFT2}), we conclude that the initial perturbations exhibit linear growth with time rather than exponential growth, indicating only the weak instability of the found trajectory.

For the time-varying sink-vortex (\ref{eq:sink+rotation}), the solution of Zermelo's equation (\ref{eq:angle}) is given by $\theta_t = \theta_0 + t/2 - t^2/2$. Then, the equation of motion (\ref{eq:motion}) can be integrated from the starting point $\bm R_0 = \bm r_A$
\begin{eqnarray}
    \nonumber
    X_t = &\dfrac{e^{-0.3 t}}{3}  \Bigg[ 3 x_A \cos \left(\dfrac{t(t-1)}{2} \right)  + 3 y_A \sin \left( \dfrac{t(t-1)}{2} \right)\\
    \label{eq:SRF1}
    &+ 10 V_0 (e^{0.3 t}-1) \cos \left( \dfrac{t(t-1)}{2} - \theta_0 \right) \Bigg],
 \end{eqnarray}
 \begin{eqnarray}
    \nonumber
    Y_t = &\dfrac{e^{-0.3 t}}{3}  \Bigg[ 3 y_A \cos \left(\dfrac{t(t-1)}{2} \right) - 3 x_A  \sin \left( \dfrac{t(t-1)}{2} \right) \\
    \label{eq:SRF2}
    & - 10 V_0 (e^{0.3 t}-1) \sin \left( \dfrac{t(t-1)}{2} - \theta_0 \right) \Bigg].
\end{eqnarray}
The found trajectory should pass through the terminal point $\bm R_T = \bm r_B$, which allows us to obtain the initial heading $\theta_0 \approx -0.78$ and travel time $T \approx 1.03$. Figure~\ref{fig:2}b shows the optimal trajectory and illustrates its robustness to small errors in the initial heading $\delta \theta_0 = \pi/360$ and position $\Delta = 0.025$. Based on expressions (\ref{eq:SRF1}) and (\ref{eq:SRF2}), we conclude that the initial perturbations do not exhibit growth with time, indicating the neutral stability of the optimal trajectory.

The remaining examples are more complex and permit only numerical solutions. The optimal trajectory obtained for the Taylor-Green flow (\ref{eq:TG-flow}) is quite expected, see Fig.~\ref{fig:2}c. For the 2D turbulence snapshot and the time-dependent flow, the optimal control equations admit many solutions, reflecting the existence of multiple local extrema in the optimization landscape (see also Ref.~\cite{biferale2019zermelo}). For illustration, we select the trajectories that yield the shortest travel time among all optimal trajectories identified in our study; see Figs.~\ref{fig:2}d and ~\ref{fig:2}e for the P1 set of starting and ending points (and the SM~\cite{SM} for the P2 set). While we cannot prove that these solutions correspond to the global optimum, they nevertheless provide upper bounds on the minimum travel time, and further we use them as benchmarks to evaluate the performance of reinforcement learning algorithms. The travel times corresponding to the identified optimal trajectories are summarized in Table~\ref{tab:1}.

\begin{figure*}[t]
\includegraphics[width=0.9\linewidth]{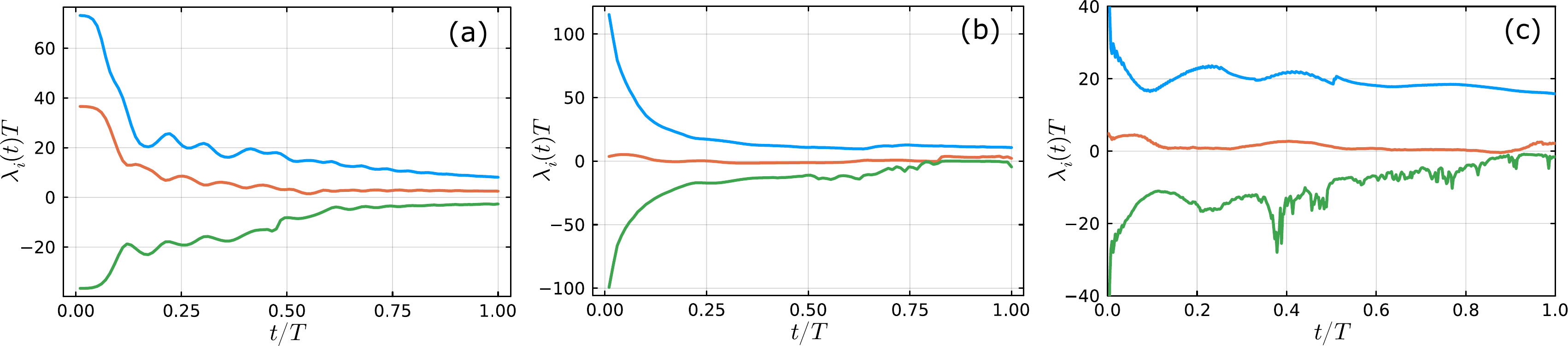}
\caption{Normalized FTLEs $\lambda_i T$ for the optimal trajectories found in (a) Taylor-Green flow, (b) turbulent snapshot, and (c) time-dependent turbulent flow, see Figs.~\ref{fig:2}c-\ref{fig:2}e.}\label{fig:3}
\end{figure*}

The main drawback of the optimal control approach is the instability of the resulting trajectories in complex environments. This instability has been previously discussed for 2D turbulence snapshots~\cite{biferale2019zermelo} and steady regular currents~\cite{piro2022efficiency, piro2021optimal}. Notably, in Ref.~\cite{piro2021optimal} it was demonstrated that initially nearby trajectories can separate rapidly after passing through centers of vortices and sinks. Therefore, these flow structures can effectively amplify small perturbations and lead to pronounced divergence in the resulting paths. 

We complement these findings and extend the analysis to time-dependent flows. First, we illustrate the instability by plotting near-optimal trajectories perturbed by small deviations in the initial heading $\delta \theta_0 = \pi/9000$ and position $\Delta = 10^{-3}$. Although these perturbations are considerably smaller than those examined in steady vortex and time-dependent sink-vortex examples, the resulting trajectories fail to reach the target $\bm r_B$, as shown in Figs.~\ref{fig:2}c-\ref{fig:2}e. Qualitatively, we find that rotational and positional uncertainties produce similar effects on trajectory divergence.

To make a more quantitative stability analysis of the optimal trajectories, we follow Ref.~\cite{biferale2019zermelo} and consider the evolution of infinitesimal perturbations,
\begin{equation}
(\delta X_t, \delta Y_t, \delta \theta_t)^{T} = \mathbb{W}(t) (\delta X_0, \delta Y_0, \delta \theta_0)^T,
\end{equation}
where $\mathbb{W}(t)$ is the linearized evolution matrix. A polar decomposition $\mathbb{W}(t) = \mathbb{V}(t)\mathbb{R}(t)$ allows us to separate the effects of rotation, represented by the orthogonal matrix $\mathbb{R}(t)$, from those of stretching, characterized by the positive-definite matrix $\mathbb{V}(t)$. The eigenvalues $h_i(t)$ of $\mathbb{V}(t)$ (equal to singular values of $\mathbb{W}(t)$) quantify the magnitude of stretching along different directions in phase space and are used to compute the finite-time Lyapunov exponents (FTLEs), defined as
\begin{equation}
\lambda_i(t) = \dfrac{1}{t} \ln h_i(t), \quad i = 1,2,3.
\end{equation}
The corresponding results are presented in Fig.~\ref{fig:3}. The presence of a positive maximal FTLE indicates that the optimal trajectories are unstable with respect to small perturbations on the time scales needed to reach the target. This instability limits the practical applicability of the pre-computed optimal control strategy, as minor deviations in initial conditions lead to significant divergence of trajectories. 

Finally, to better understand the instability, it is instructive to contrast it with the behavior of tracers in steady two-dimensional incompressible flows. For such tracers, motion occurs strictly along the streamlines, which constrains their trajectories to be regular and precludes the onset of chaotic dynamics. In the present case, however, the agent possesses self-propulsion and an additional degree of freedom associated with the steering direction $\theta_t$, which breaks the integrability of the purely passive system. As a result, the considered active system becomes far more susceptible to chaotic behavior even in steady incompressible flows.

\section{Reinforcement Learning}\label{sec:5}

In this section, we use reinforcement learning (RL) to develop stable navigation strategies. We focus on two RL approaches: Q-learning and a one-step actor-critic algorithm. After a brief introduction, we apply these methods to the presented examples and compare their performance with optimal-control solutions. For a comprehensive overview of the RL algorithms used here, we refer the reader to Ref.~\cite{sutton1998reinforcement}. 

The present analysis is largely motivated by the study of the one-step actor-critic algorithm in Ref.~\cite{biferale2019zermelo}. Building on that framework, we additionally incorporate Q-learning and extend the comparison between optimal-control solutions and RL-based predictions to include time-dependent external flows and a broader set of environments. A related study~\cite{nasiri2022reinforcement} investigates the more advanced A2C algorithm; however, it focuses on steady currents and considers different flow configurations.

\subsection{Overview of the Algorithms}

To introduce RL framework, we consider an agent that moves between states $s$ by performing actions $a$. At each step the agent selects an action randomly with a probability given by a policy function $\pi(a|s)$ and receives a reward $r$. A trajectory can therefore be written as $s_0, a_0, r_1, s_1, \dots, s_{T-1}, a_{T-1}, r_T, s_T$, where $T$ denotes the end of an episode. The expected cumulative reward obtained when starting from state $s$ and following policy $\pi$ defines the state-value function
\begin{equation}
v_{\pi} (s) = \mathbb{E}_{\pi} \left[r_{t+1} +  \dots + r_T \mid s_t=s \right],
\end{equation}
while the action-value function
\begin{equation}
q_{\pi} (s,a) = \mathbb{E}_{\pi} \left[r_{t+1} + \dots + r_T \mid s_t=s, a_t=a \right]
\end{equation}
gives the expected cumulative reward when the agent takes action $a$ in state $s$.

The objective is to determine an optimal policy $\pi^*(a|s)$ that maximizes the expected cumulative reward. Action-value methods focus on learning the corresponding function $q_{*}(s,a)$, which satisfies the Bellman optimality equation. In Q-learning, $q_{*}(s,a)$ is estimated iteratively through interaction with the environment using the update rule:
\begin{equation}\label{eq:Q-update}
\begin{aligned}
Q(s_t,a_t) &\leftarrow Q(s_t,a_t) + \alpha \Big[\, r_{t+1} \\
&\quad + \max_{a'} Q(s_{t+1},a') - Q(s_t,a_t) \,\Big],
\end{aligned}
\end{equation}
where $\alpha$ is the learning rate and $Q(s,a)$ is the current approximation of the optimal action-value function. Over many iterations, provided all state-action pairs are adequately explored, $Q(s,a)$ converges to $q_{*}(s,a)$. The optimal policy is then obtained by selecting, in each state, the action that maximizes the action-value function $\pi^*= \argmax_a q_{*}(s,a)$. A key strength of Q-learning is its off-policy nature: the optimal policy can be learned while following a different exploratory strategy. To balance exploration and exploitation during training, we employ an $\varepsilon$-greedy protocol, in which the action maximizing $Q(s,a)$ is chosen with probability $1-\varepsilon$, while a random action is selected with probability $\varepsilon$. 

Next, we consider the actor-critic algorithm, a policy-gradient method combining two components: an actor, representing the parameterized policy $\pi(a|s,\bm p)$, and a critic, which estimates the state-value function via $V(s,\mathbf{w})$. Both sets of parameters $\bm p$ and $\bm w$ are iteratively updated using gradient-based rules to converge toward the optimal policy $\pi^*(a|s)$ and value function $v_{*}(s)$. In the one-step formulation, the updates read:
\begin{eqnarray}
    \label{eq:update_1}
    \bm p_{t+1} = \bm p_t &+& \alpha_p \Big[ r_{t+1} + V(s_{t+1}, \bm w_t)\\
    \nonumber
    &-&V(s_{t}, \bm w_t) \Big] \nabla_{\bm p} \ln \pi(a_t|s_t,\bm p_t),\\
    \label{eq:update_2}
    \bm w_{t+1} = \bm w_t &+& \alpha_{w} \Big[ r_{t+1} + V(s_{t+1}, \bm w_t)\\
    \nonumber
    &-&V(s_{t}, \bm w_t) \Big] \nabla_{\bm w} V(s_t, \bm w_t),
\end{eqnarray}
where $\alpha_p$ and $\alpha_w$ are the actor and critic learning rates. The actor learns to favor actions that produce higher-than-expected rewards, guided by the advantage $r_{t+1} + V(s_{t+1}, \bm w_t) - V(s_{t}, \bm w_t)$, while the critic improves its estimate of expected cumulative rewards using the same signal. Note that once training is complete, we evaluate the learned strategy in a testing phase by selecting the greedy actions $a_t =\argmax_a \pi(a|s_t, \bm p)$. This deterministic choice allows us to highlight the most likely trajectories implied by the learned policy.  

\begin{figure*}[t]
\includegraphics[width=0.9\linewidth]{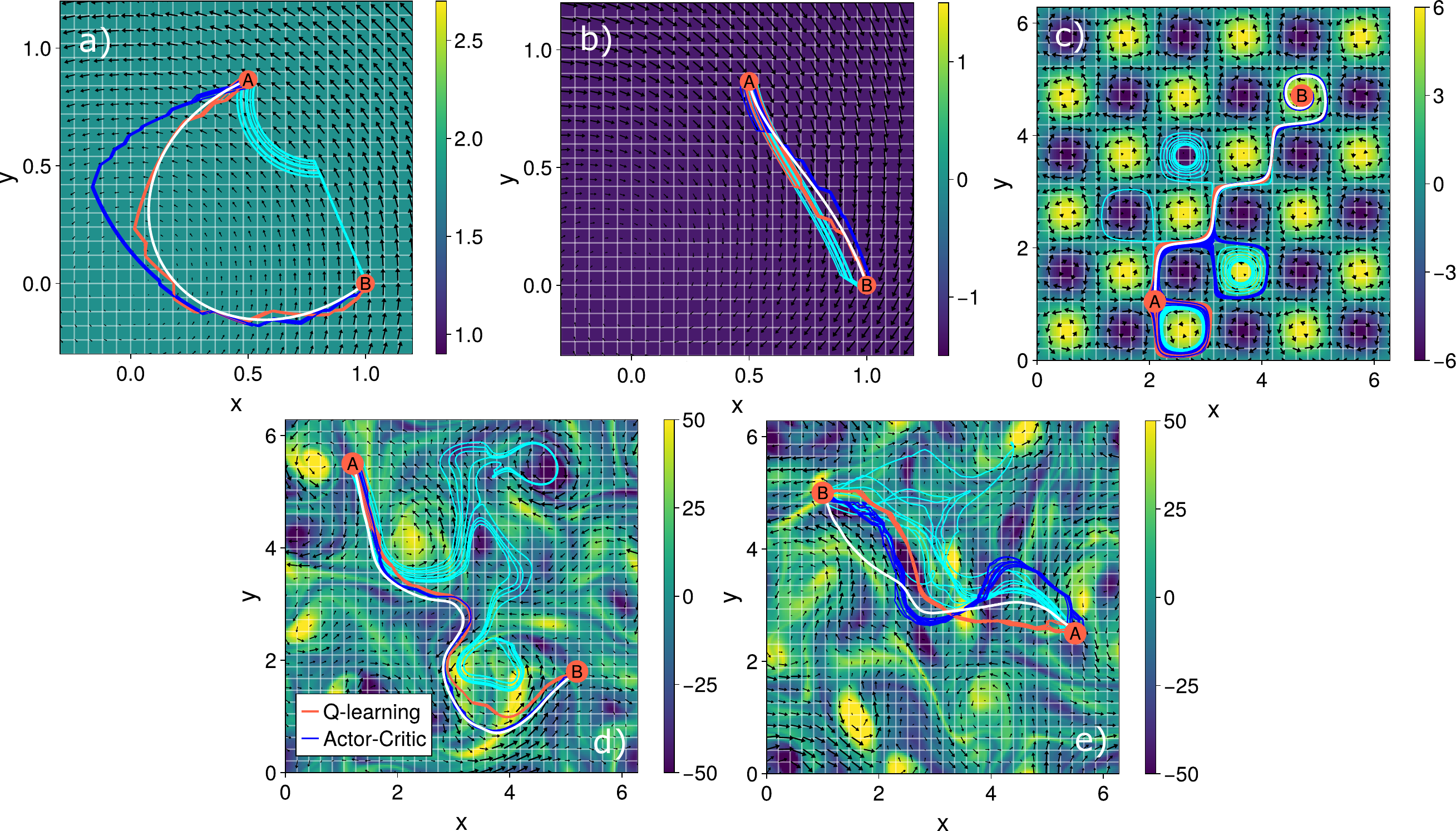}
\caption{Trajectories generated by Q-learning (red) and one-step actor-critic (blue) agents, shown alongside the optimal-control solutions (white) and trivial policy (cyan) for (a) steady vortex, (b) time-varying sink-vortex, (c) Taylor-Green flow, (d) turbulent snapshot (P1), and (e) time-dependent turbulent flow (P1).}\label{fig:4}
\end{figure*}

\subsection{States, Actions, Rewards}

In Zermelo's problem, we discretize the spatial domain $L \times L$ into square cells of size $\delta \times \delta$, with $\delta = L/N_{tile}$, yielding $N_s = N_{tile}^2$ states. Roughly speaking, this coarse graining implies that the agent resolves its position up to the cell size $\delta$. Steering decisions are taken at intervals $\Delta t=\delta/(2u_0)$, where $u_0$ is of the order of the maximum flow velocity. The action space is also discretized: control variable $\theta_t$ can take one of $N_a$ equally spaced angles $\theta_j = -\pi + 2 \pi j/N_a$ with $j=1,2,\dots,N_a$. During each step of duration $\Delta t$, the trajectory is advanced by numerically integrating Eq.~(\ref{eq:motion}) with fixed steering.

An episode corresponds to one attempt to reach the target and terminates when the agent enters a circle of radius $d_B$ centered at $\bm r_B$, when the maximum time $T_{\max}$ is exceeded, or upon collision with the domain boundary. Boundary collisions incur an additional negative reward $-2 T_{max}$, providing a strong penalty. The reward function favors short travel times,
\begin{equation}
    r_t = -\Delta t + \dfrac{|\bm r_B - \bm R_{t-\Delta t}|}{V_0} - \dfrac{|\bm r_B - \bm R_{t}|}{V_0},
\end{equation}
where $V_0$ is the slip velocity and $\bm R_t$ is the agent position. The first term penalizes travel time, while the remaining terms measure progress toward the target and are known to preserve optimality while significantly accelerating learning~\cite{biferale2019zermelo, gunnarson2021learning}. When summed over a successful episode (the agent has reached the target), they add an approximately constant offset $\approx |\bm r_B-\bm r_A|/V_0$. To improve robustness to initial-position perturbations, $\bm R_0$ is sampled uniformly from a circle of radius $d_A$ centered at $\bm r_A$ at the start of each episode. Throughout this work, we set $d_A = d_B \approx \delta \ll |\bm r_B - \bm r_A|$.

In the above formulation, the system state is defined solely by the agent's position. While sufficient for steady flows, this representation may limit performance in time-dependent flows since the optimal action at a given location can vary in time. Nevertheless, for the time-dependent scenarios examined in this work, we found that RL algorithms employing this reduced state representation can still cope with temporal variability and learn control strategies that reach the target.

With discrete states $i=1,\dots,N_s$ and actions $j=1,\dots,N_a$, the Q-learning action-value function is represented by a matrix $Q_{ij}$. For the actor-critic algorithm, we adopt the softmax policy parametrization
\begin{equation}
\pi(a_j| s_i, \bm p) = \dfrac{\exp (p_{ij})}{\sum_{k=1}^{N_a} \exp (p_{ik})},
\end{equation}
where $p_{ij}$ are trainable parameters. The state-value function is parameterized as $V(s_i,\bm w)=w_i$, assigning one parameter per state.

\subsection{Training Hyperparameters}

The domain sizes correspond to those shown in Fig.~\ref{fig:2}. The spatial and action discretizations are chosen similarly to Ref.~\cite{biferale2019zermelo}: we set $N_{tile}=25-30$ and $N_a=8$. The maximum episode duration $T_{max}$ is selected to be several times larger than the typical travel time. This allows the agent sufficient time to explore the environment, while also serving to terminate training episodes in which the agent becomes trapped. For Q-learning, we use a constant learning rate $\alpha$. The $\varepsilon$-greedy policy exploration parameter is annealed from $1$ to $0.01$ by multiplying it by a constant factor after each episode. This factor is chosen such that $\varepsilon=0.01$ is reached after approximately half of the total number of training episodes. In the one-step actor-critic algorithm, we also employ constant learning rates for the actor $\alpha_p$ and critic $\alpha_w$, with $\alpha_w = 2 \alpha_p$. We found that the training process is sensitive to the specific choice of learning rates; therefore, these parameters were tuned separately for each example. With this tuning, stable training is achieved: the learning curve initially increases and then reaches a plateau, and this behavior is consistently reproduced across independent training runs. Compared with Ref.~\cite{biferale2019zermelo}, we use a larger number of training episodes, $n_{train} = 2 \times 10^4 - 10^5$, together with relatively small learning rates. After training, the learned strategies are evaluated over $n_{eval} = 5 \times 10^3$ episodes. The specific values of all hyperparameters used in each example are reported in the SM~\cite{SM}.

\subsection{Results}

We find that for all considered background flows both RL algorithms successfully solve the navigation problem, i.e., the resulting trajectories connect the starting and ending points, see Fig.~\ref{fig:4} (and the SM~\cite{SM} for the P2 set of points in turbulent examples). This result is robust: its reproducibility was verified in five independent training runs for regular flows and in ten independent runs for turbulent flows. During the evaluation stage, more than 99\% of trajectories successfully reach the target in every example. Therefore, the main parameter characterizing the efficiency of the RL solutions is the mean travel time of successful trajectories. This quantity fluctuates across independent training runs, and we report the corresponding mean values and standard deviations in Table~\ref{tab:1}. For comparison, we also implement a trivial policy, where the steering angle selected by the agent at each $\Delta t$ is given by the action that points most directly toward the target among the $N_a=8$ different possibilities. The resulting trajectories are shown in Fig.~\ref{fig:4} and we discuss them below.

\begin{table}[b]
\caption{Travel times for optimal-control trajectories ($T$), Q-learning ($T_Q$), and actor-critic algorithm ($T_{AC}$). "S" corresponds to steady flows and "T" to time-dependent flows.}
\begin{ruledtabular}
\begin{tabular}{c|c|c|c}
 & $T$ & $\langle T_Q \rangle$ & $\langle T_{AC} \rangle$\\
\hline
vortex (S)        & 1.97 & $2.17 \pm 0.05$ & $2.19 \pm 0.01$\\
sink-vortex (T)   & 1.03 & $1.06 \pm 0.01$ & $1.09 \pm 0.02$\\
TG flow (S)       & 12.2 & $13.0 \pm 0.2$  & $13.1 \pm 0.5$\\
2D turb (S), P1   & 0.82 & $0.94 \pm 0.03$ & $0.87 \pm 0.005$\\
2D turb (T), P1   & 0.67 & $0.90 \pm 0.06$ & $1.17 \pm 0.04$\\
2D turb (S), P2   & 0.82 & $0.99 \pm 0.05$ & $1.03 \pm 0.01$\\
2D turb (T), P2   & 0.70 & $1.04 \pm 0.05$ & $1.00 \pm 0.03$\\
\end{tabular}
\end{ruledtabular}
\label{tab:1}
\end{table}

For the steady vortex, the actor-critic method produces trajectories that systematically deviate from the optimal path, see Fig.~\ref{fig:4}a. These deviations are consistent across independent training runs and persist even when the training hyperparameters are varied. Q-learning sometimes yields trajectories that closely follow the optimal-control solution, see Fig.~\ref{fig:4}a, although noticeable deviations can also occur. Despite these differences, the corresponding travel times are similar, suggesting that the minimum of the policy landscape is relatively flat. The performance of both RL algorithms remains within about 10\% of the optimal-control time, see Table~\ref{tab:1}. The trivial policy also succeeds in solving the navigation problem in this case, but its travel time $\langle T_{TP} \rangle \approx 7.5$ is significantly larger, as the agent moves against the strong flow. 

For the time-varying sink-vortex, both RL algorithms produce qualitatively similar trajectories that fluctuate around the optimal-control solution, see Fig.~\ref{fig:4}b. The mean travel time obtained with Q-learning is slightly lower and lies within 3\% of the optimal-control time, whereas the actor-critic result is within 6\%, see Table~\ref{tab:1}. However, this example is relatively simple and can also be solved by the trivial policy, which outperforms both RL methods and yields a travel time of $\langle T_{TP} \rangle \approx 1.04$.

Next, we consider the Taylor-Green flow. The resulting trajectories are generally close to the optimal-control path, but they may include a few additional revolutions within nearby vortices, see Fig.~\ref{fig:4}c. Both RL methods yield comparable mean travel times, exceeding the optimal-control time by about 7\%. The trivial policy solves the problem only with probability $\approx 0.3$, depending on the starting position of the agent. In the remaining trials, the agent becomes trapped in Taylor-Green vortices, from which this naive strategy cannot escape. The mean travel time of successful attempts is $\langle T_{TP} \rangle \approx 11.1$ and it is shorter than the optimal-control prediction due to the finite sizes of the starting and target regions, which are similar to those used in the RL framework.

For the turbulence snapshot, the actor-critic agent outperforms the Q-learning agent for the P1 set of starting and ending points, and vice versa for the P2 set, see Table~\ref{tab:1}. The found trajectories are close to the optimal-control paths in terms of how they exploit the surrounding flow, see Fig.~\ref{fig:4}d. The mean travel times exceed the optimal-control solutions by approximately 6-25\%, depending on the RL algorithm and the specific set of points considered. The trivial policy fails completely for both sets of points. Under this strategy, the agent eventually enters region where the external flow cancels its self-propulsion, causing it to stall. As a result, it becomes immobilized and is unable to reach the target.

Finally, for the time-dependent turbulent flow, Q-learning and the actor-critic method converge to different policies, which is reflected in the distinct trajectories followed by the trained agents, see Fig.~\ref{fig:4}e. However, neither of these routes is close to the optimal-control prediction, and the performance of the RL solutions deteriorates noticeably: the mean travel time exceeds the optimal-control result by 35–75\%, see Table~\ref{tab:1}. This behavior is likely related to the reduced state-space representation, which relies solely on the agent's position and neglects the time-dependence of the flow. These findings suggest that there is room for improvement through more advanced navigation algorithms. The trivial policy solves the navigation problem for the P1 set of points with probability $\approx 0.96$, but yields an even longer mean travel time $\langle T_{TP} \rangle \approx 1.41$, and it almost completely fails for the P2 set of points~\cite{SM}.

Additionally, we conducted training with a finer discretization of the state-action space, using $N_{tile}=40-50$ and $N_a=12$. Qualitatively, the learned strategies lead to similar trajectories. Quantitatively, the mean travel time remains approximately the same or slightly decreases for regular flows (with the exception of the one-step actor-critic algorithm and the Taylor-Green flow, where the improvement is noticeable) and remains the same or increases for turbulent flows. The increase of travel time for complex turbulent settings is related to convergence of RL algorithms to local minima. For tabular RL methods, the likelihood of such sub-optimal convergence increases as the dimensionality of the phase space grows, since the enlarged state-action space makes efficient exploration and accurate value estimation more challenging. This suggests that the performance of RL methods in turbulent settings is not primarily limited by state-space discretization, but rather by the ability of the algorithms to converge to global optima. Further details are provided in the SM~\cite{SM}.

\section{Incomplete Flow Information}\label{sec:6}

So far, we have shown that RL algorithms successfully solve navigation problems in various background flows and remain robust to perturbations in the agent's initial position. We now investigate their robustness to incomplete flow information in turbulent environments, a situation relevant to route planning in atmospheric or oceanic currents where forecasts are usually coarse-grained. An alternative approach for navigation in such settings is to improve the quality of forecasts, e.g., using super-resolution techniques~\cite{fukami2023super, sofos2025review, parfenyev2024inferring}. We focus on the stability of RL-based strategies under incomplete flow information and therefore we do not apply such methods.

\begin{figure*}[t]
\includegraphics[width=\linewidth]{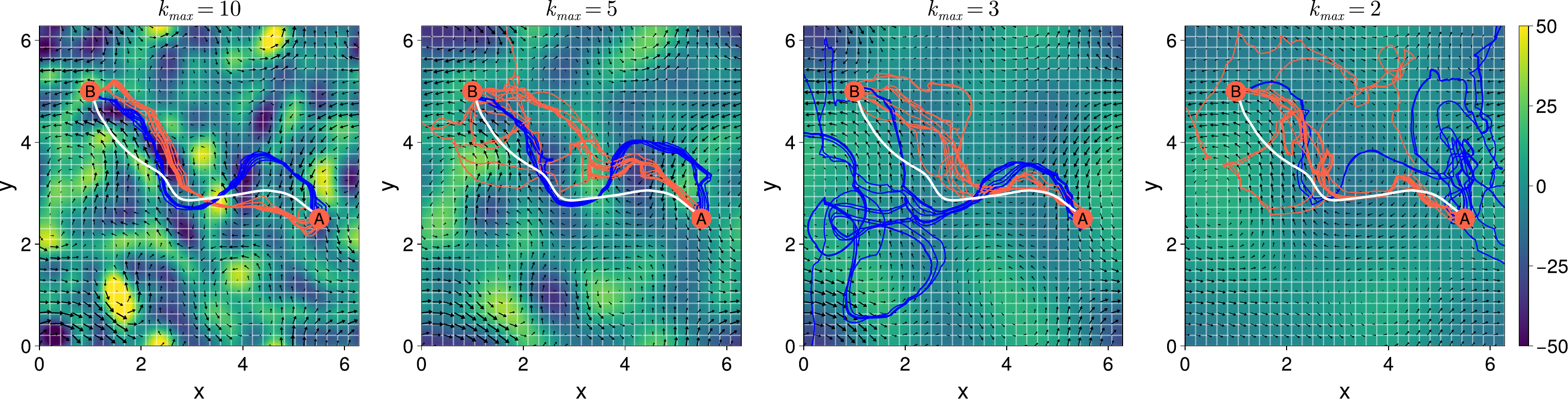}
\caption{Trajectories generated by the Q-learning (red) and one-step actor-critic (blue) agents are shown alongside the optimal-control path (white) for the time-dependent turbulent flow (P1). The background illustrates the coarse-grained velocity field employed during training, while all trajectories were evaluated in the full velocity field.}\label{fig:5}
\end{figure*}

We assume that the RL agent has access only to flow modes with wavenumbers $k \leq k_{max}$, while smaller-scale fluctuations are unresolved. Practically, we filter the DNS velocity field by removing modes with $k > k_{max}$ and train the agent on this coarse-grained flow. The learned strategy is then evaluated in the full velocity field to assess whether the agent can still reach the target. We examine two scenarios: navigation in a steady turbulence snapshot and in a fully time-dependent turbulent flow. In each case, we analyze two pairs of starting and ending points, labeled P1 and P2. The turbulence forcing scale is fixed at $k_f=5.5$, while the filtering scale is varied across $k_{max} = \{256, 10, 5, 3, 2 \}$, with $k_{max} = 256$ representing the fully resolved DNS field. Training hyperparameters are identical to those used earlier. Performance is evaluated in terms of both the probability of successfully reaching the target and the mean travel time of successful trajectories. All results are averaged over ten independent training runs to reduce stochastic variability in the training process.

Representative trajectories of RL agents trained in coarse-grained flows are illustrated in Fig.~\ref{fig:5}. When the filtering scale is relatively small, the trajectories are close to those obtained in the full velocity field. As the filtering scale increases, unresolved velocity fluctuations begin to noticeably distort the trajectories, and RL paths starting from slightly different initial positions can diverge significantly. In addition, the fraction of trajectories that reach the target decreases.

\begin{figure}[b]
\includegraphics[width=0.8\linewidth]{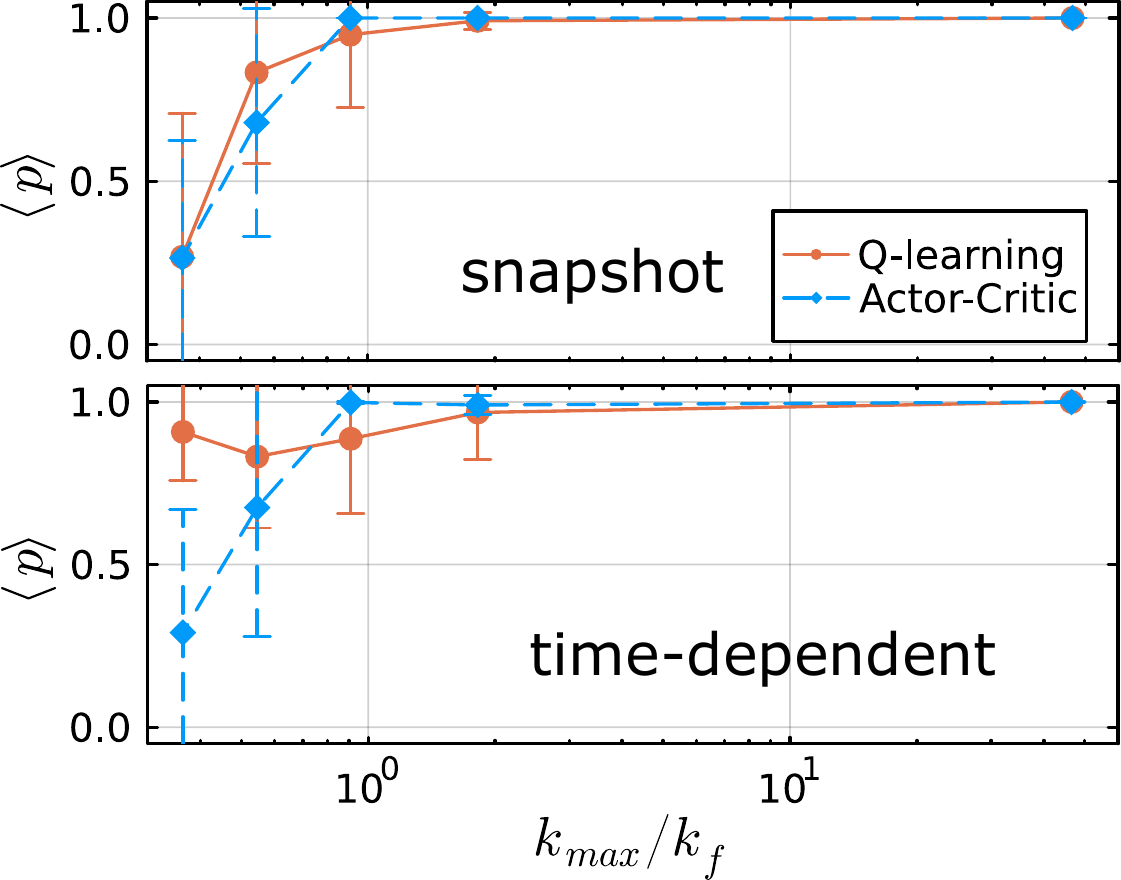}
\caption{Success rate vs. normalized filtering scale $k_{max}/k_f$ for the turbulence snapshot (top) and time-dependent turbulent flow (bottom). Error bars show the standard deviation.}\label{fig:6}
\end{figure}

We find that the results obtained in the coarse-grained flows are extremely noisy and highly sensitive to the choice of starting and ending points. For example, in a turbulence snapshot with $k_{max}=2$, the Q-learning agent fails to reach the target for the P1 set of points, whereas the actor-critic agent achieves a success probability close to $0.5$. However, for the P2 set of points, the probabilities remain approximately the same, but the algorithms switch roles. To identify overall trends, Fig.~\ref{fig:6} shows how the probability of reaching the target depends on the filtering scale. The data are averaged not only over training runs but also over the P1 and P2 sets of points. The higher Q-learning success rate at $k_{max}=2$ in the time-dependent flow is related to the P2 points, where the target-reaching probability rises to near unity. This likely stems from a coincidental alignment of conditions and appears specific to the chosen initial and target locations.

\begin{figure}[b]
\includegraphics[width=0.8\linewidth]{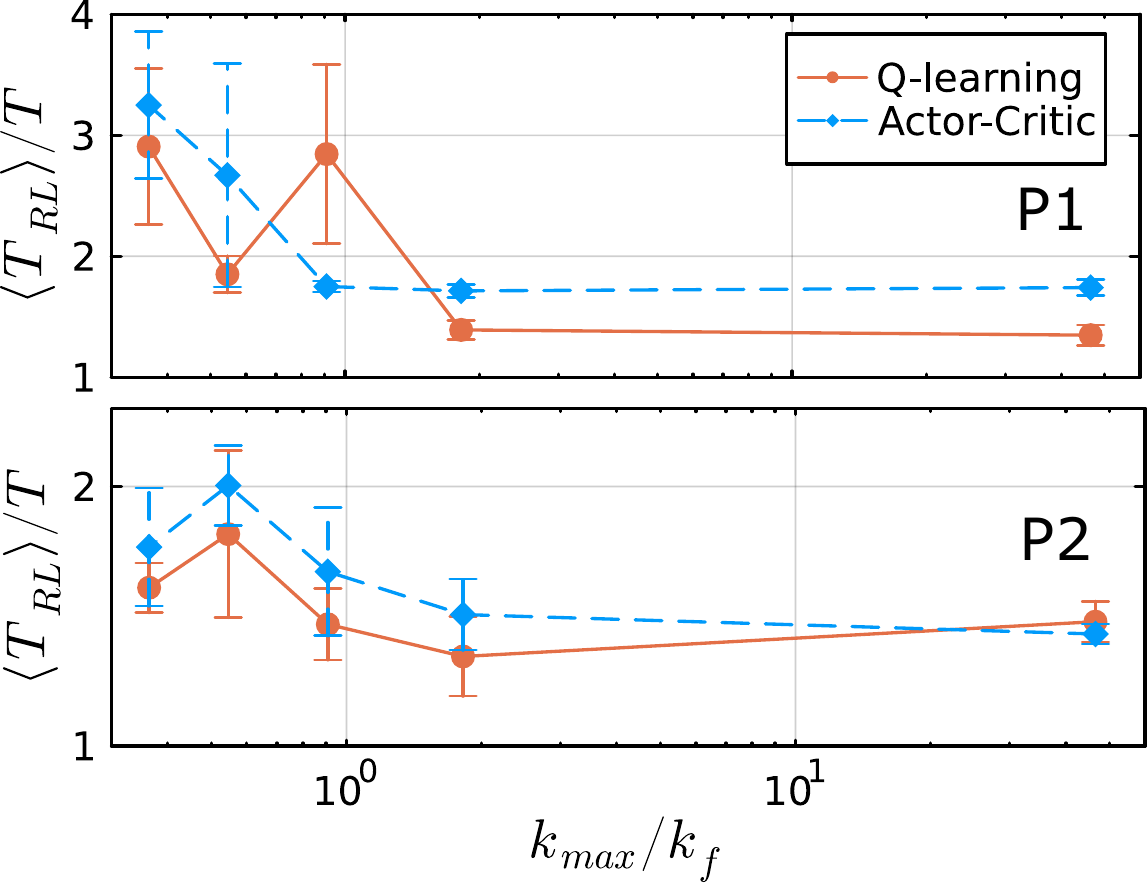}
\caption{Mean travel time of successful trajectories, normalized by the optimal-control time $T$, as a function of the normalized filtering scale $k_{max}/k_f$ for the time-dependent turbulent flow. The top panel corresponds to the P1 set of points, and the bottom panel to the P2 set.}\label{fig:7}
\end{figure}

In turbulent flows, most of the energy is concentrated at large scales, see Fig.~\ref{fig:1}, and therefore these scales primarily determine the route chosen by the agent. Small-scale fluctuations have relatively low velocities, and as long as they remain small compared to the agent's self-propulsion speed $V_0$, the agent can easily overcome their influence. As the filtering scale increases (i.e., as $k_{max}$ decreases), the characteristic velocity of the unresolved scales grows, and this noise begins to noticeably affect the motion, see Fig.~\ref{fig:5}. Qualitatively, this leads to a reduction in the probability of reaching the target, as observed in Fig.~\ref{fig:6}. As the success probability decreases, we also observe an overall trend toward an increase in the mean travel time of successful trajectories, reflecting changes in their shape. Although somewhat noisy, this trend is illustrated by the dependencies shown in Fig.~\ref{fig:7} for time-dependent turbulent flows. These results suggest that the growing influence of unresolved turbulent fluctuations not only lowers the probability of reaching the target but also reduces the efficiency of successful navigation.

\section{Conclusion}\label{sec:7}

In this work, we systematically investigated optimal navigation in two-dimensional flows of increasing dynamical complexity, from steady solid-body rotation and time-dependent sink-vortex to Taylor-Green flow and turbulence in the inverse energy cascade regime. Using control theory, we showed that locally optimal time-minimizing trajectories can be computed even in complex time-dependent environments. However, these trajectories are often unstable and highly sensitive to initial conditions, which limits their practical use. Nevertheless, they provide valuable reference solutions and upper bounds for the minimum travel time.

As a more robust alternative, we explored reinforcement learning (RL). We implemented Q-learning and a one-step actor-critic algorithm and demonstrated that both methods reliably reach the target across all considered flow configurations, with failure rates close to zero. While the trajectories produced by the two methods may differ, neither algorithm consistently outperforms the other. Instead, their performance appears complementary depending on the flow configuration.

The travel times achieved by RL agents exceed the optimal-control solutions by about 3–10\% in regular flows, while the gap increases to 35–75\% in time-dependent turbulent environments. This indicates substantial room for improvement, suggesting that more advanced deep RL approaches~\cite{mecanna2025critical} could further reduce this discrepancy. Initial progress in this direction has been reported in Ref.~\cite{nasiri2022reinforcement}, where the A2C method reproduced analytical optimal-control solutions for simple steady flows, while time-dependent configurations remained outside the scope of that study.

Finally, we examined the generalization capability of agents trained on coarse-grained turbulent velocity fields. Despite having access only to large-scale flow information, the agents successfully navigated in the fully resolved turbulent environment. This robustness to missing small-scale information is particularly encouraging for real-world applications, where flow forecasts are typically low-resolution and uncertain. Importantly, the success of RL in this setting appears to rely primarily on the large-scale physical structure of the flow rather than on algorithmic complexity. Similar behavior may therefore be expected in three-dimensional turbulence, where the energy distribution across scales follows a comparable Kolmogorov spectrum, $E(k) \propto k^{-5/3}$, but the question requires further investigation.\\

\acknowledgments

The work was supported by the Ministry of Science and Higher Education of the Russian Federation (within the state assignment of the Landau Institute for Theoretical Physics) and by the Basic research program of HSE University. I am also grateful to the Landau Institute for providing computing resources.

\section*{DATA AVAILABILITY}

The data that support the findings of this study are available from the corresponding author upon reasonable request.

\bibliography{biblio}

@article{zermelo1931navigationsproblem,
  title={{\"U}ber das Navigationsproblem bei ruhender oder ver{\"a}nderlicher Windverteilung},
  author={Zermelo, Ernst},
  journal={Z. Angew. Math. Mech.},
  volume={11},
  number={2},
  pages={114--124},
  year={1931},
  publisher={Wiley Online Library}
}

@article{nelson2010microrobots,
  title={Microrobots for minimally invasive medicine},
  author={Nelson, Bradley J and Kaliakatsos, Ioannis K and Abbott, Jake J},
  journal={Annual review of biomedical engineering},
  volume={12},
  number={1},
  pages={55--85},
  year={2010},
  publisher={Annual Reviews}
}

@article{park2017multifunctional,
  title={Multifunctional bacteria-driven microswimmers for targeted active drug delivery},
  author={Park, Byung-Wook and Zhuang, Jiang and Yasa, Oncay and Sitti, Metin},
  journal={ACS nano},
  volume={11},
  number={9},
  pages={8910--8923},
  year={2017},
  publisher={ACS Publications}
}

@article{manzari2021targeted,
  title={Targeted drug delivery strategies for precision medicines},
  author={Manzari, Mandana T and Shamay, Yosi and Kiguchi, Hiroto and Rosen, Neal and Scaltriti, Maurizio and Heller, Daniel A},
  journal={Nature Reviews Materials},
  volume={6},
  number={4},
  pages={351--370},
  year={2021},
  publisher={Nature Publishing Group UK London}
}

@inproceedings{trincavelli2008towards,
  title={Towards environmental monitoring with mobile robots},
  author={Trincavelli, Marco and Reggente, Matteo and Coradeschi, Silvia and Loutfi, Amy and Ishida, Hiroshi and Lilienthal, Achim J},
  booktitle={2008 IEEE/RSJ International Conference on Intelligent Robots and Systems},
  pages={2210--2215},
  year={2008},
  organization={IEEE}
}

@article{lermusiaux2017future,
  title={A future for intelligent autonomous ocean observing systems},
  author={Lermusiaux, Pierre FJ and Subramani, DN and Lin, J and Kulkarni, CS and Gupta, A and Dutt, A and Lolla, T and Haley Jr, PJ and Ali, W Hajj and Mirabito, C and others},
  journal={Journal of Marine Research},
  volume={75},
  pages={765–813},
  year={2017}
}

@article{bellemare2020autonomous,
  title={Autonomous navigation of stratospheric balloons using reinforcement learning},
  author={Bellemare, Marc G and Candido, Salvatore and Castro, Pablo Samuel and Gong, Jun and Machado, Marlos C and Moitra, Subhodeep and Ponda, Sameera S and Wang, Ziyu},
  journal={Nature},
  volume={588},
  number={7836},
  pages={77--82},
  year={2020},
  publisher={Nature Publishing Group UK London}
}

@article{chai2020monitoring,
  title={Monitoring ocean biogeochemistry with autonomous platforms},
  author={Chai, Fei and Johnson, Kenneth S and Claustre, Herv{\'e} and Xing, Xiaogang and Wang, Yuntao and Boss, Emmanuel and Riser, Stephen and Fennel, Katja and Schofield, Oscar and Sutton, Adrienne},
  journal={Nature Reviews Earth \& Environment},
  volume={1},
  number={6},
  pages={315--326},
  year={2020},
  publisher={Nature Publishing Group UK London}
}

@book{bryson2018applied,
  title={Applied Optimal Control: Optimization, Estimation, and Control},
  author={Bryson, Arthur Earl and Ho Yu-Chi},
  year={2018},
  publisher={Routledge}
}

@article{techy2011optimal,
  title={Optimal navigation in planar time-varying flow: Zermelo’s problem revisited},
  author={Techy, Laszlo},
  journal={Intelligent Service Robotics},
  volume={4},
  number={4},
  pages={271--283},
  year={2011},
  publisher={Springer}
}

@book{kirk2004optimal,
  title={Optimal Control Theory: An Introduction},
  author={Kirk, Donald E},
  year={2004},
  publisher={Courier Corporation}
}

@book{vallis2017atmospheric,
  title={Atmospheric and Oceanic Fluid Dynamics},
  author={Vallis, Geoffrey K},
  year={2017},
  publisher={Cambridge University Press}
}

@book{angelo2009chaos,
  title={Chaos: from simple models to complex systems},
  author={Angelo, Vulpiani and Fabio, Cecconi and Massimo, Cencini},
  year={2009},
  publisher={World Scientific Publishing Company}
}

@article{piro2022optimal,
  title={Optimal navigation of microswimmers in complex and noisy environments},
  author={Piro, Lorenzo and Mahault, Beno{\^\i}t and Golestanian, Ramin},
  journal={New Journal of Physics},
  volume={24},
  number={9},
  pages={093037},
  year={2022},
  publisher={IOP Publishing}
}

@article{monthiller2022surfing,
  title={Surfing on turbulence: a strategy for planktonic navigation},
  author={Monthiller, R{\'e}mi and Loisy, Aurore and Koehl, Mimi AR and Favier, Benjamin and Eloy, Christophe},
  journal={Physical Review Letters},
  volume={129},
  number={6},
  pages={064502},
  year={2022},
  publisher={APS}
}

@article{calascibetta2023optimal,
  title={Optimal tracking strategies in a turbulent flow},
  author={Calascibetta, Chiara and Biferale, Luca and Borra, Francesco and Celani, Antonio and Cencini, Massimo},
  journal={Communications Physics},
  volume={6},
  number={1},
  pages={256},
  year={2023},
  publisher={Nature Publishing Group UK London}
}

@article{yoo2016path,
  title={Path optimization for marine vehicles in ocean currents using reinforcement learning},
  author={Yoo, Byunghyun and Kim, Jinwhan},
  journal={Journal of Marine Science and Technology},
  volume={21},
  number={2},
  pages={334--343},
  year={2016},
  publisher={Springer}
}

@article{colabrese2017flow,
  title={Flow navigation by smart microswimmers via reinforcement learning},
  author={Colabrese, Simona and Gustavsson, Kristian and Celani, Antonio and Biferale, Luca},
  journal={Physical Review Letters},
  volume={118},
  number={15},
  pages={158004},
  year={2017},
  publisher={APS}
}

@article{gustavsson2017finding,
  title={Finding efficient swimming strategies in a three-dimensional chaotic flow by reinforcement learning},
  author={Gustavsson, Kristian and Biferale, Luca and Celani, Antonio and Colabrese, Simona},
  journal={The European Physical Journal E},
  volume={40},
  number={12},
  pages={110},
  year={2017},
  publisher={Springer}
}

@article{schneider2019optimal,
  title={Optimal steering of a smart active particle},
  author={Schneider, Elias and Stark, Holger},
  journal={Europhysics Letters},
  volume={127},
  number={6},
  pages={64003},
  year={2019},
  publisher={IOP Publishing}
}

@article{biferale2019zermelo,
  title={Zermelo’s problem: optimal point-to-point navigation in 2D turbulent flows using reinforcement learning},
  author={Biferale, Luca and Bonaccorso, Fabio and Buzzicotti, Michele and Clark Di Leoni, Patricio and Gustavsson, Kristian},
  journal={Chaos: An Interdisciplinary Journal of Nonlinear Science},
  volume={29},
  number={10},
  year={2019},
  publisher={AIP Publishing}
}

@inproceedings{buzzicotti2020optimal,
  title={Optimal control of point-to-point navigation in turbulent time dependent flows using reinforcement learning},
  author={Buzzicotti, Michele and Biferale, Luca and Bonaccorso, Fabio and Clark di Leoni, Patricio and Gustavsson, Kristian},
  booktitle={International Conference of the Italian Association for Artificial Intelligence},
  pages={223--234},
  year={2020},
  organization={Springer}
}

@article{alageshan2020machine,
  title={Machine learning strategies for path-planning microswimmers in turbulent flows},
  author={Alageshan, Jaya Kumar and Verma, Akhilesh Kumar and Bec, J{\'e}r{\'e}mie and Pandit, Rahul},
  journal={Physical Review E},
  volume={101},
  number={4},
  pages={043110},
  year={2020},
  publisher={APS}
}

@article{gunnarson2021learning,
  title={Learning efficient navigation in vortical flow fields},
  author={Gunnarson, Peter and Mandralis, Ioannis and Novati, Guido and Koumoutsakos, Petros and Dabiri, John O},
  journal={Nature Communications},
  volume={12},
  number={1},
  pages={7143},
  year={2021},
  publisher={Nature Publishing Group UK London}
}

@article{nasiri2023optimal,
  title={Optimal active particle navigation meets machine learning (a)},
  author={Nasiri, Mahdi and L{\"o}wen, Hartmut and Liebchen, Benno},
  journal={Europhysics Letters},
  volume={142},
  number={1},
  pages={17001},
  year={2023},
  publisher={IOP Publishing}
}

@book{sutton1998reinforcement,
  author = {Sutton, Richard S. and Barto, Andrew G.},
  edition = {Second},
  publisher = {The MIT Press},
  title = {Reinforcement Learning: An Introduction},
  year = {2018 }
}

@article{boffetta2012two,
  title={Two-dimensional turbulence},
  author={Boffetta, Guido and Ecke, Robert E},
  journal={Annual Review of Fluid Mechanics},
  volume={44},
  number={1},
  pages={427--451},
  year={2012},
  publisher={Annual Reviews}
}

@article{filatov2016nonlinear,
  title={Nonlinear generation of vorticity by surface waves},
  author={Filatov, S. V. and Parfenyev, V. M. and Vergeles, S. S. and Brazhnikov, M. Yu. and Levchenko, A. A. and Lebedev, V. V.},
  journal={Physical Review Letters},
  volume={116},
  number={5},
  pages={054501},
  year={2016},
  publisher={APS},
  language={english}
}

@article{filatov2016generation,
  title={Generation of vortices by gravity waves on a water surface},
  author={Filatov, S. V. and Aliev, S. A. and Levchenko, A. A. and Khramov, D. A.},
  journal={JETP Letters},
  volume={104},
  number={10},
  pages={702--708},
  year={2016},
  publisher={Springer},
  language={english}
}

@article{francois2017wave,
  title={Wave-based liquid-interface metamaterials},
  author={Francois, N. and Xia, H. and Punzmann, H. and Fontana, P. W. and Shats, M.},
  journal={Nature Communications},
  volume={8},
  number={1},
  pages={1--9},
  year={2017},
  publisher={Nature Publishing Group},
  language={english}
}

@article{xia2019generation,
  title={Generation of vortex lattices at the liquid--gas interface using rotating surface waves},
  author={Xia, Hua and Francois, Nicolas and Gorce, Jean-Baptiste and Punzmann, Horst and Shats, Michael},
  journal={Fluids},
  volume={4},
  number={2},
  pages={74},
  year={2019},
  publisher={MDPI},
  language={english}
}

@article{abella2020measurement,
  title={Measurement of Eulerian vorticity beneath rotating surface waves},
  author={Abella, Alfred P and Soriano, Maricor N},
  journal={Physica Scripta},
  volume={95},
  number={8},
  pages={085007},
  year={2020},
  publisher={IOP Publishing}
}

@article{GeophysicalFlowsJOSS,
  year = {2021},
  publisher = {The Open Journal},
  volume = {6},
  number = {60},
  pages = {3053},
  author = {Navid C. Constantinou and Gregory LeClaire Wagner and Lia Siegelman and Brodie C. Pearson and André Palóczy},
  title = {{GeophysicalFlows.jl: Solvers for geophysical fluid dynamics problems in periodic domains on CPUs \& GPUs}},
  journal = {Journal of Open Source Software}
}

@article{valadao2025spectrum,
  title={Spectrum correction in Ekman-Navier-Stokes turbulence},
  author={Valad{\~a}o, VJ and Boffetta, G and De Lillo, F and Musacchio, S and Crialesi-Esposito, M},
  journal={Journal of Turbulence},
  pages={1--10},
  year={2025},
  publisher={Taylor \& Francis}
}

@article{von2011double,
  title={Double cascade turbulence and Richardson dispersion in a horizontal fluid flow induced by Faraday waves},
  author={{Von Kameke}, A. and Huhn, F. and Fern{\'a}ndez-Garc{\'\i}a, G. and Munuzuri, A. P. and P{\'e}rez-Mu{\~n}uzuri, V.},
  journal={Physical Review Letters},
  volume={107},
  number={7},
  pages={074502},
  year={2011},
  publisher={APS},
  language={english}
}

@article{francois2013inverse,
  title={{Inverse energy cascade and emergence of large coherent vortices in turbulence driven by Faraday waves}},
  author={Francois, N. and Xia, H. and Punzmann, H. and Shats, M.},
  journal={Physical Review Letters},
  volume={110},
  number={19},
  pages={194501},
  year={2013},
  publisher={APS},
  language={english}
}

@article{francois2014three,
  title={{Three-dimensional fluid motion in Faraday waves: creation of vorticity and generation of two-dimensional turbulence}},
  author={Francois, N. and Xia, H. and Punzmann, H. and Ramsden, S. and Shats, M.},
  journal={Physical Review X},
  volume={4},
  number={2},
  pages={021021},
  year={2014},
  publisher={APS},
  language={english}
}

@article{colombi2021three,
  title={Three dimensional flows beneath a thin layer of 2D turbulence induced by Faraday waves},
  author={Colombi, Raffaele and Schl{\"u}ter, Michael and {von Kameke}, Alexandra},
  journal={Experiments in Fluids},
  volume={62},
  pages={1--13},
  year={2021},
  publisher={Springer}
}

@article{colombi2022coexistence,
  title={Coexistence of inverse and direct energy cascades in faraday waves},
  author={Colombi, Raffaele and Rohde, Niclas and Schl{\"u}ter, Michael and {von Kameke}, Alexandra},
  journal={Fluids},
  volume={7},
  number={5},
  pages={148},
  year={2022},
  publisher={MDPI}
}

@article{sommeria1986experimental,
  title={Experimental study of the two-dimensional inverse energy cascade in a square box},
  author={Sommeria, Joel},
  journal={Journal of Fluid Mechanics},
  volume={170},
  pages={139--168},
  year={1986},
  publisher={Cambridge University Press}
}

@article{paret1998intermittency,
  title={Intermittency in the two-dimensional inverse cascade of energy: Experimental observations},
  author={Paret, Jerome and Tabeling, Patrick},
  journal={Physics of Fluids},
  volume={10},
  number={12},
  pages={3126--3136},
  year={1998},
  publisher={American Institute of Physics}
}

@article{xia2009spectrally,
  title={Spectrally condensed turbulence in thin layers},
  author={Xia, H and Shats, M and Falkovich, Gregory},
  journal={Physics of Fluids},
  volume={21},
  number={12},
  pages={125101},
  year={2009},
  publisher={American Institute of Physics}
}

@article{bardoczi2014experimental,
  title={Experimental confirmation of self-regulating turbulence paradigm in two-dimensional spectral condensation},
  author={Bard{\'o}czi, L and Bencze, A and Berta, M and Schmitz, L},
  journal={Physical Review E},
  volume={90},
  number={6},
  pages={063103},
  year={2014},
  publisher={APS}
}

@article{orlov2018large,
  title={Large-scale coherent vortex formation in two-dimensional turbulence},
  author={Orlov, Artur Valer'evich and Brazhnikov, M Yu and Levchenko, Aleksandr Alekseevich},
  journal={JETP Letters},
  volume={107},
  pages={157--162},
  year={2018},
  publisher={Springer}
}

@article{zhu2024flow,
  title={Flow patterns and energy spectra in forced quasi-two-dimensional turbulence: effect of system size and damping rate},
  author={Zhu, Hang-Yu and Xie, Jin-Han and Xia, Ke-Qing},
  journal={Journal of Fluid Mechanics},
  volume={996},
  pages={A39},
  year={2024},
  publisher={Cambridge University Press}
}

@article{piro2022efficiency,
  title={Efficiency of navigation strategies for active particles in rugged landscapes},
  author={Piro, Lorenzo and Golestanian, Ramin and Mahault, Beno{\^\i}t},
  journal={Frontiers in Physics},
  volume={10},
  pages={1034267},
  year={2022},
  publisher={Frontiers Media SA}
}

@article{piro2021optimal,
  title={Optimal navigation strategies for microswimmers on curved manifolds},
  author={Piro, Lorenzo and Tang, Evelyn and Golestanian, Ramin},
  journal={Physical Review Research},
  volume={3},
  number={2},
  pages={023125},
  year={2021},
  publisher={APS}
}

@article{liebchen2019optimal,
  title={Optimal navigation strategies for active particles},
  author={Liebchen, Benno and L{\"o}wen, Hartmut},
  journal={Europhysics Letters},
  volume={127},
  number={3},
  pages={34003},
  year={2019},
  publisher={EDP Sciences, IOP Publishing and Societ{\`a} Italiana di Fisica}
}

@article{nasiri2022reinforcement,
  title={Reinforcement learning of optimal active particle navigation},
  author={Nasiri, Mahdi and Liebchen, Benno},
  journal={New Journal of Physics},
  volume={24},
  number={7},
  pages={073042},
  year={2022},
  publisher={IOP Publishing}
}

@article{mecanna2025critical,
  title={A critical assessment of reinforcement learning methods for microswimmer navigation in complex flows: S. Mecanna et al.},
  author={Mecanna, Selim and Loisy, Aurore and Eloy, Christophe},
  journal={The European Physical Journal E},
  volume={48},
  number={10},
  pages={58},
  year={2025},
  publisher={Springer}
}

@article{yang2020efficient,
  title={Efficient navigation of colloidal robots in an unknown environment via deep reinforcement learning},
  author={Yang, Yuguang and Bevan, Michael A and Li, Bo},
  journal={Advanced Intelligent Systems},
  volume={2},
  number={1},
  pages={1900106},
  year={2020},
  publisher={Wiley Online Library}
}

@article{muinos2021reinforcement,
  title={Reinforcement learning with artificial microswimmers},
  author={Muinos-Landin, Santiago and Fischer, Alexander and Holubec, Viktor and Cichos, Frank},
  journal={Science Robotics},
  volume={6},
  number={52},
  pages={eabd9285},
  year={2021},
  publisher={American Association for the Advancement of Science}
}

@article{fukami2023super,
  title={Super-resolution analysis via machine learning: a survey for fluid flows: K. Fukami et al.},
  author={Fukami, Kai and Fukagata, Koji and Taira, Kunihiko},
  journal={Theoretical and Computational Fluid Dynamics},
  volume={37},
  number={4},
  pages={421--444},
  year={2023},
  publisher={Springer}
}

@article{sofos2025review,
  title={A review of deep learning for super-resolution in fluid flows},
  author={Sofos, Filippos and Drikakis, Dimitris},
  journal={Physics of Fluids},
  volume={37},
  number={4},
  year={2025},
  publisher={AIP Publishing}
}

@article{parfenyev2024inferring,
  title={Inferring parameters and reconstruction of two-dimensional turbulent flows with physics-informed neural networks},
  author={Parfenyev, Vladimir and Blumenau, Mark and Nikitin, Ilia},
  journal={JETP Letters},
  volume={120},
  number={8},
  pages={599--607},
  year={2024},
  publisher={Springer}
}

@unpublished{SM,
  author={{See the Supplemental Material at [URL will be inserted by publisher] for additional details and illustrations.}}
}

@article{parfenyev2019formation,
  title={Formation and decay of eddy currents generated by crossed surface waves},
  author={Parfenyev, VM and Filatov, SV and Brazhnikov, M Yu and Vergeles, SS and Levchenko, AA},
  journal={Physical Review Fluids},
  volume={4},
  number={11},
  pages={114701},
  year={2019}
}

\clearpage 
\onecolumngrid 
\pagestyle{empty} 
\setlength{\hoffset}{-1in}
\setlength{\voffset}{-1in}
\setlength{\topmargin}{0pt}
\setlength{\headheight}{0pt}
\setlength{\headsep}{0pt}
\setlength{\oddsidemargin}{0pt}
\noindent\includegraphics[page=1,width=\paperwidth,height=\paperheight,keepaspectratio]{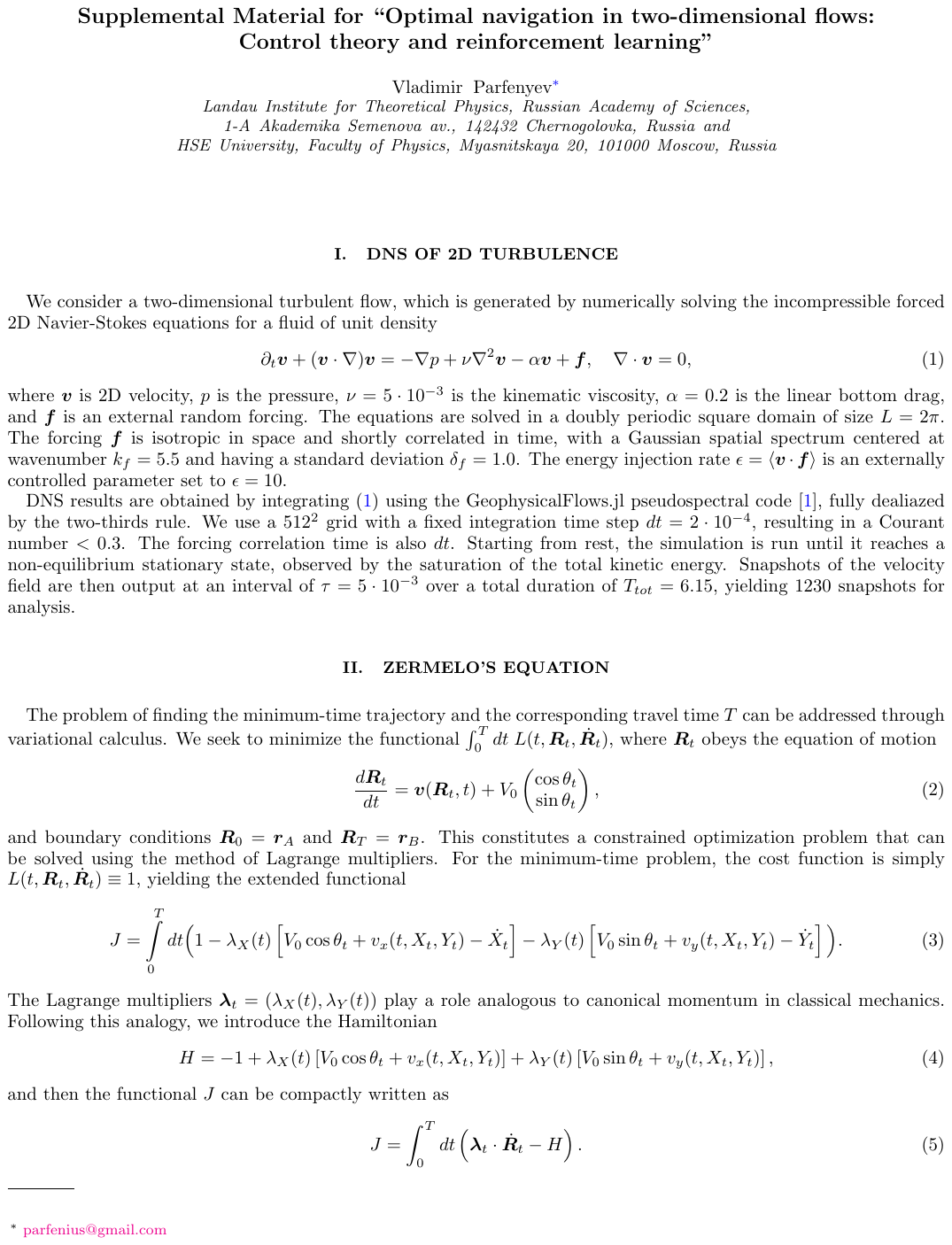}
\clearpage
\noindent\includegraphics[page=2,width=\paperwidth,height=\paperheight,keepaspectratio]{SM.pdf}
\clearpage
\noindent\includegraphics[page=3,width=\paperwidth,height=\paperheight,keepaspectratio]{SM.pdf}
\clearpage
\noindent\includegraphics[page=4,width=\paperwidth,height=\paperheight,keepaspectratio]{SM.pdf}
\clearpage
\noindent\includegraphics[page=5,width=\paperwidth,height=\paperheight,keepaspectratio]{SM.pdf}
\clearpage

\end{document}